\documentstyle[epsf,prb,floats,aps]{revtex}
\begin{document}
\draft
\title  {First-principles and semi-empirical calculations 
        for bound hole polarons in KNbO$_3$}
\author {E.~A.~Kotomin and R.~I.~Eglitis}
\address{Universit\"at Osnabr\"uck -- Fachbereich Physik, 
         D-49069 Osnabr\"uck, Germany,\\
	 and
         Institute of Solid State Physics, University of Latvia, 
	 8 Kengaraga, Riga LV-1063, Latvia}
\author {A.~V.~Postnikov and G.~Borstel}
\address{Universit\"at Osnabr\"uck -- Fachbereich Physik, 
         D-49069 Osnabr\"uck, Germany}
\author {N.~E.~Christensen}
\address{Institute of Physics and Astronomy, University of Aarhus,
         Aarhus C, DK-8000, Denmark}
\date{March 24, 1999}
\twocolumn[\hsize\textwidth\columnwidth\hsize\csname@twocolumnfalse\endcsname
\maketitle

\begin{abstract}
The {\it ab initio} linear muffin-tin-orbital (LMTO) formalism and
the semi-empirical method of the Intermediate Neglect of the
Differential Overlap (INDO) based on the Hartree--Fock formalism
are combined for the study of the hole polarons (a hole trapped
nearby the cation vacancy) in a cubic phase of KNbO$_3$ perovskite
crystals. The 40-atom and 320-atom supercells were used,
respectively. We predict existence of both, one-site and two-site
(molecular) polarons with close optical absorption energies (0.9
eV and 0.95 eV). The relevant experimental data are discussed.
\end{abstract}
\pacs{PACS numbers:
  77.84.Dy,     71.15.Fv,      71.10.+x,     77.80.Bh    }
]

\section{Introduction}
It is well understood now that {\it point defects} play an
important role in the electro-optical and non-linear optical
applications of KNbO$_3$ and related ferroelectric materials.\cite{gun1} 
In particular, reduced KNbO$_3$ crystals containing
oxygen vacancies reveal fast photorefractive response to short-pulse 
excitations which could be
used for developing fast optical correlators.\cite{zgo} 
The prospects of the use of KNbO$_3$ for the light frequency doubling 
are seriously affected by the presence of unidentified defects 
responsible for induced infrared absorption.\cite{polzik} 
The photorefractive effect, important in particular
for holographic storage, is also well known to depend on the
presence of impurities and defects. Most of as-grown ABO$_3$
perovskite crystals are non-stoichiometric and thus contain
considerable amount of vacancies.

The so-called $F$ and $F^+$ centers 
(an O vacancy, V$_{\mbox{\tiny O}}$, which traps 
two or one electron, respectively),\cite{chen,hughes,kot3} 
belong to the most common defects in oxide crystals.
In electron-irradiated KNbO$_3$, a broad absorption band observed
around 2.7 eV at room temperature has been tentatively
ascribed to the $F$-type centers.\cite{hod} 
These two defects were the subject of our recent {\it ab initio} and
semi-empirical calculations.\cite{F+,F} 
A transient optical absorption band at 1.2 eV has been associated
recently,\cite{gri} in analogy with other perovskites, with a hole
polaron (a hole bound, probably, to a K vacancy). The ESR study of KNbO$_3$
doped with Ti$^{4+}$ gives a proof that holes could be trapped by
such negatively charged defects.\cite{poss} 
For example, in BaTiO$_3$, the hole polarons bound to Na and
K alkali ions replacing Ba and thus forming a negatively charged
site attracting a hole,\cite{var} have also been found.
Cation vacancies are the most likely candidates for pinning 
polarons. In irradiated MgO, they are known to
trap one or two holes giving rise to the V$^-$ and V$^0$
centers\cite{chen,hughes} which are nothing but bound hole polaron
and bipolaron, respectively. The results of the experimental studies
of hole polarons in alkali halides and ferroelectric perovskites
reveal two different forms of atomic structure of polarons:
atomic one (one-site), when a hole is localized on a single atom, and
molecular-type (two-site), when a hole is shared by two atoms
forming a quasi-molecule.\cite{poss,var,stone} 
In the present study, we simulate hole polarons
associated with a K vacancy in KNbO$_3$, using an {\it ab initio}
density functional theory (DFT)-based method in combination with
a semi-empirical treatment based on the Hartree-Fock (HF) formalism,
employing the periodic boundary conditions and the supercell geometry 
in both cases.

\section{Methods}\label{sec:methods}

The motivation for using the DFT-based
and HF-based calculation methods in parallel
is to combine strong sides of both in a single study.
The DFT is expected to be able to provide good description of the
ground state, i.e. to deliver reasonable relaxation energies
and ground-state geometry. In the HF approach, the relaxation
energies are generally less accurate because of
the neglection of correlation effects.
On the other hand, the HF formalism is straightforwardly
suited for the evaluation of excitation energies, because the
total energies can be calculated for any (ground-state or excited)
electronic configuration on equal footing, that is
generally not the case in the DFT.
Practical experience shows that HF and DFT results
often exhibit similar qualitative trends in the description
of dielectric properties but quantitatively 
lie on opposite sides of experimental data, thus effectively
setting error bars for a theoretical prediction.\cite{HF-Res}

Our {\it ab initio} DFT treatment is based on the 
full-potential linearized muffin-tin orbital (LMTO) formalism,
previously applied with success to the study of structural instability 
and lattice dynamics in pure KNbO$_3$.\cite{ktn3,phonon} 
For the study of defects, we used the version of LMTO as implemented
by van Schilfgaarde and Methfessel\cite{fplmto}, that was earlier used 
in our simulations for the $F$-center in KNbO$_3$ (Ref.~\onlinecite{F}; 
see also for more details of the calculation setup there). 
The exchange-correlation has been treated 
in the local density approximation (LDA), as parametrized
by Perdew and Zunger.\cite{perzu}
The supercell LMTO approach has been earlier successfully used 
for the simulation of defects in KCl\cite{chr1} and MgO.\cite{chr2} 
In the present case we used the
$2\!\times\!2\!\times\!2$ supercells, i.e. the distance 
between repeated point defects was $\approx$8 \AA.
As a consequence of the large number of eigenstates per {\bf k}-point
in a reduced Brillouin zone (BZ) of the supercell and
of the metallicity of the doped system, it was essential
to maintain a quite dense mesh for the {\bf k}-integration
by the tetrahedron method over the BZ. Specifically, clear
trends in the total energy as function of atomic displacements
were only established at $10\!\times\!10\!\times\!10$ divisions
of the BZ (i.e., 186 irreducible {\bf k}-points for a one-site polaron).

\begin{figure}[b]
\epsfxsize=8.5cm \centerline{\epsfbox{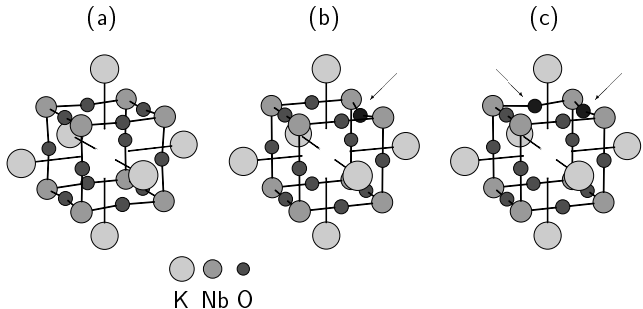}} %\vspace*{0.5cm}
\caption{Uniform relaxation of O atoms around the K vacancy in KNbO$_3$
(a) and the relaxation corresponding to the formation of one-site (b)
and two-site (c) polarons. Arrows in (b) and (c)
indicate the displaced O atoms.}
\label{fig:struc}
\end{figure}

In the HF formalism, a semi-empirical 
Intermediate Neglect of the Differential Overlap (INDO)\cite{indo} method,
modified for ionic and partly ionic solids,\cite{indo1,indo2} has been used.
The INDO method has been successfully applied for the simulation
of defects in many oxides.\cite{F+,F,kot2,kot1,sta1,rob}
The calculations have been performed with periodic boundary
conditions in the so-called large unit cell (LUC) model,\cite{evar}
i.e., for ${\bf k}$=0 in the appropriately reduced BZ.
When mapped on the conventional BZ of the compound in question, 
this accounts for the band dispersion effects and allows to
incorporate effectively the ${\bf k}$-summation 
on a relatively fine mesh over the BZ. Due to the robustness
of the summation procedure, the total energy dependence
on the atomic displacements was found satisfactorily smooth
for any supercell size (one should keep in mind, however, that
the numerical results may remain somehow dependent on the LUC in question).
In the present work, $4\!\times\!4\!\times\!4$ supercells (320 atoms)
were used, that is the extension of our preliminary 
INDO study\cite{egl3} with the $2\!\times\!2\!\times\!2$ supercells.
The detailed analysis of the INDO parametrisation 
for KNbO$_3$ is presented in Ref.~\onlinecite{egl1},
and the application of the method to the $F$-center calculations --
in Ref.~\onlinecite{F+,F}.

\begin{figure}[t]
\epsfxsize=7.8cm \centerline{\epsfbox{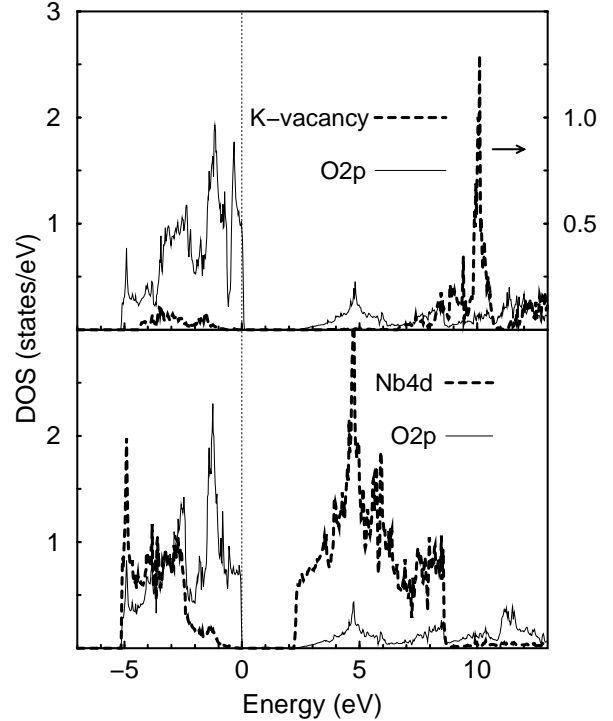}} %\vspace*{0.8cm}
\caption{Local DOS at the K vacancy site and at the adjacent
oxygen atom (top panel) and at Nb and O sites in perfect KNbO$_3$
(bottom panel), as calculated by LMTO.}
\label{fig:DOS}
\end{figure}

We restricted ourselves to a cubic phase of KNbO$_3$, 
with the lattice constant $a_0$=4.016~{\AA}. In the vacancy-containing
supercell, the relaxation of either one (for the one-site
polaron) or two neighboring (for the two-site polaron) O atoms,
amongst twelve closest to the K vacancy, has been allowed for,
and the changes in the total energy (as compared to the
unrelaxed perovskite structure with a K atom removed)
have been analyzed. Also, we studied
the fully symmetric relaxation pattern (breathing of twelve
O atoms) around the vacancy.
Different relaxation patterns considered in the present study
are shown in Fig.~\ref{fig:struc}. The positions of more distant
atoms in the supercell were kept fixed.

\section{Results}

The removal from the supercell of a K atom with its 7 electrons 
contributing to the valence band (VB)
produces slightly different effects on the electronic
structure, as described within the DFT and in the HF
formalism. Acording to the LMTO result, the Fermi energy lowers, 
and the system
becomes metallic (remaining non-magnetic). Therefore, no specific
{\it occupied} localized state is associated with the vacancy.
The local density of states (DOS) at the sites of interest
is shown in Fig.~\ref{fig:DOS}.
As is typical for LDA calculations, the one-electron band gap in KNbO$_3$
comes out underestimated ($\approx$2 eV) as compared to the
experimental optical gap ($\approx$3.3 eV). 
The removal of a K$4s$ electron amounts to adding a hole which
forms a localized state at $\approx$10 eV
above the Fermi level, i.e. above the unoccupied Nb$4d$ band.
In the $2p$-DOS of O atoms neighboring the vacancy, a quasi-local
state (that effectively screens the hole)
is visible just below the Fermi level. Apart from that,
the O$2p$-DOS is largely unaffected by the presence of vacancy,
and the changes in the DOS of more distant sites (K, Nb) are
negligible as compared with those in the perfect crystal.
As the cubic symmetry is lifted by allowing a non-uniform relaxation
of O atoms, the ``screening'' quasi-local state is clearly
localized at the atom closest to the vacancy. At the same time,
the hole state becomes smeared out in energy. This amounts to
the bonding being established between the hole and the screening
charge on one of its neighbors.

\begin{figure}[b]
\epsfxsize=6.0cm \centerline{\epsfbox{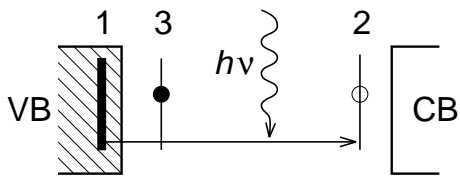}} \vspace*{0.4cm}
\caption{Sketch of the polaron optical transition from the quasi-local state
{\it 1} near the top of the valence band to the hole state {\it 2} below the
conduction band bottom. {\it 3} indicates the level of an unpaired electron.}
\label{fig:HF_band}
\end{figure}

In the INDO treatment, the one-electron optical gap is overestimated,
as is typical for the HF calculations 
($\approx$6 eV, see Ref.~\onlinecite{egl1}),
but the $\Delta$SCF gap for the triplet state is 2.9 eV,
close to the experiment.
The quasi-local ``screening'' state is described by 
a wide band close to the VB top. 
This is consistent with the LDA description.
But the INDO calculation also suggests, and this differs from the LDA, that
the removal of an electron
leaves an unpaired electron state split-off at $\approx$1eV 
above the VB band top.
In case of asymmetrical O relaxation, the 
molecular orbital associated with this state is centered
at the displaced O atom, being a combination of the $2p_x$, $2p_y$ functions
of the latter in the setting when the plane spanned
by their lobes passes through the vacancy site.
The same applies qualitatively to the two-site polaron, 
with the only difference
that the localized state is formed from the $2p$ orbitals of {\it both}
O atoms approaching the vacancy, with a corresponding symmetry lowering.
The localized hole state is also present in the HF description
but lies much lower than the corresponding state in the LDA,
forming a 0.9 eV -wide band located $\approx$ 0.2 eV 
below the conduction band bottom (see Fig.~\ref{fig:HF_band}).

\begin{figure}[t]
\epsfxsize=8.8cm \centerline{\epsfbox{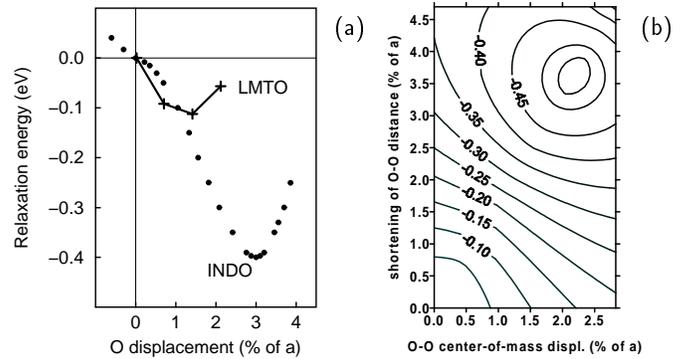}} %\vspace*{0.6cm}
\caption{
Relaxation energy for one-site polaron according to LMTO and INDO
calculations vs the displacement of one oxygen atom towards the vacancy (a);
relaxation energy for two-site polaron as function of the center-of-mass
displacement of the O-O pair and of the O-O distance
according to the INDO calculation (b).
}
\label{fig:DE}
\end{figure}

Differently from the DFT-based approaches
which address in principle only the ground-state electron density,
the HF method provides a possibility to evaluate excitation
energies by means of the so-called $\Delta$SCF formalism,
i.e. as the difference of total energies in relaxed ground-state
and excited states. 
In agreement with the Schirmer's theory for the small-radius
polarons in ionic solids,\cite{schirm} the optical absorption
corresponds to a hole transfer to the state delocalized over
nearest oxygens. The absorption energies due to the electron
transition from the quasi-local states near the VB top ({\it 1}, 
Fig.~\ref{fig:HF_band})
into the vacant polaron band ({\it 2}, Fig.~\ref{fig:HF_band})
for one-site and two-site polarons are close (Table I),
and both are twice smaller than the experimental value 
for a hole polaron trapped by the Ti impurity.\cite{poss} 
This shows that the optical absorption energy of small bound polarons 
can be strongly dependent on the defect involved. 
Another important observation is that the $\Delta$SCF energy
for absorption turns out to be considerably smaller than the estimate
based on the difference of one-electron energies.

\begin{table}[b]
\caption{Absorption ($E_{\mbox{\small abs}}$) and lattice
relaxation ($E_{\mbox{\small rel}}$,
relatively to the perfect crystal with a K vacancy) 
energies (eV), calculated by LMTO and INDO methods.}
\begin{tabular}{dddd}
Energy:           &E$_{abs}$ &E$_{rel}$ &E$_{rel}$ \\
Method:           & INDO & LMTO & INDO\\
\hline
uniform breathing &      & 0.01 & 0.08 \\
one-site polaron  & 0.90 & 0.12 & 0.40 \\
two-site polaron  & 0.95 & 0.18 & 0.53
\end{tabular}
\end{table}

In spite of generally observed considerable degree of
covalency in KNbO$_3$ and contrary to a delocalized character of
the $F$ center state,\cite{F+,F} the one-site polaron state remains well
localized at the displaced O atom, with only a small contribution
from atomic orbitals of other O ions but none from K or Nb ions. 
Although there are some differences in the description 
of the (one-particle) electronic structure within 
the DFT- and HF-based methods, the trends in the total energy 
driving the structure optimization remain essentially the same.
In both approaches, both one-site and two-site configurations 
of the hole polaron are much more energetically favorable 
than the fully symmetric (breathing mode) relaxation 
of twelve O atoms around the K vacancy.
This is in line with what is known about small-radius polarons
in other ionic solids\cite{stone,schirm} and is caused by the fact that
the lattice polarization induced by a point charge is much larger
than that due to a delocalized charge.

In the case of one-site polaron,
a single O$^-$ ion is displaced towards the K vacancy
by 1.5 \% of the lattice constant (LMTO) or by 3\% (INDO) -- 
see Fig.~\ref{fig:DE}. The
INDO calculations show that simultaneously, 11 other nearest
oxygens surrounding the vacancy tend to be slightly displaced outwards
the vacancy. In the two-site (molecular) configuration, a hole is
shared by the two O atoms which approach each other -- by 0.5\%
(LMTO) or 3.5\% (INDO) -- and both shift towards a vacancy -- by
1.1\% (LMTO) or 2.5\% (INDO). The lattice relaxation energies
(which could be associated with the experimentally measurable hole
thermal ionization energies) are presented in Table I. In both
methods the two-site configuration of a polaron is lower in energy.

A comparison of the present, 320-atomic INDO calculation with
a preliminary calculation\cite{egl3} using 40--atomic LUC and
self--consistency in the $\Gamma$ point of the BZ only shows that
the optical absorption energies are changed inconsiderably, unlike
the lattice relaxation energies. The latter now are much smaller
and thus in better agreement with the LMTO calculation.

\section{Conclusions}
In this pilot study we focused on the quantitative models of 
hole polarons in KNbO$_3$. The main conclusion is that both
one-center and two-center configurations are energetically
favorable and close in energy (with a slight prefernce
for the two-center configuration), as follows from the numerical
simulation results by two different theoretical methods. 
The calculated optical absorption energies and the spatial
distribution of relevant electronic states
could provide guidelines for more direct experimental
identification of the defects in question. The calculated hole
polaron absorption ($\approx$1 eV) is close to the observed
short-lived absorption band energy\cite{gri}; hence this band
could indeed arise due to a hole polaron bound to a cation vacancy.
Further detailed study is needed to clarify whether such hole
polarons are responsible for the effect of the
blue light-induced infrared absorption reducing the
second-harmonic generation efficiency in KNbO$_3$.\cite{polzik}

As compared with the DFT results, the INDO (as is generally typical
for the HF-based methods) systematically gives larger
atomic displacements and relaxation energies. 
The DFT results are more reliable in what regards the ground state
of polarons, whereas the use of the HF formalism was crucial
for the calculation of their optical absorption and hence
possible experimental identification.

\acknowledgments
This study was partly supported by the DFG 
(a grant to E.~K.; the participation of A.~P. and G.~B.~in the SFB 225), 
Volkswagen Foundation (grant to R.~E.), and the
Latvian National Program on New Materials for Micro- and
Optoelectronics (E.~K.). Authors are greatly indebted 
to L.~Grigorjeva, D.~Millers, M.~R.~Philpott, A.~I.~Popov
and O.~F.~Schirmer for fruitful discussions.

\end{document}